\documentclass{svmult}
\pdfoutput=1
\usepackage[T1]{fontenc}
\usepackage[latin1]{inputenc}
\usepackage{amsmath,amssymb}
\usepackage{graphicx}
\usepackage{xspace}

\makeatletter
\def\fps@figure{tbp}
\makeatother
              
\newcommand\egaldef{\stackrel{\mbox{\upshape\tiny def}}{=}}
\newcommand\1{\leavevmode\hbox{\rm \small1\kern-0.35em\normalsize1}}
\newcommand\ind[1]{\1_{\{#1\}}}        
\newcommand\EE{\mathbb{E}}
\def\DD{\displaystyle}

\DeclareMathOperator*{\ra}{\rightarrow}
\DeclareMathOperator*{\pro}{\propto}

\newcommand\tasep{\textsc{tasep}\xspace}

\begin{document}
\title*{A queueing theory approach for a multi-speed exclusion process.}
\author{Cyril Furtlehner\inst{1} \and Jean-Marc Lasgouttes\inst{2}}

\institute{INRIA Futurs, Project Team TAO -- Bat. 490 -- Université
  Paris-Sud -- 91405 Orsay Cedex. \texttt{cyril.furtlehner@inria.fr} 
\and
  INRIA Paris-Rocquencourt, Project Team IMARA -- Domaine de Voluceau --
  BP. 105 -- 78153 Le Chesnay Cedex.
  \texttt{jean-marc.lasgouttes@inria.fr}}

\maketitle
\begin{abstract}
  We consider a one-dimensional stochastic reaction-diffusion
  generalizing the totally asymmetric simple exclusion process, and aiming
  at describing single lane roads with vehicles that can change speed.
  To each particle is associated a jump rate, and the particular
  dynamics that we choose (based on 3-sites patterns) ensures that
  clusters of occupied sites are of uniform jump rate.  When this
  model is set on a circle or an infinite line, classical arguments
  allow to map it to a linear network of queues (a zero-range process in
  theoretical physics parlance) with exponential service times, but
  with a twist: the service rate remains constant during a busy
  period, but can change at renewal events. We use the tools of
  queueing theory to compute the fundamental diagram of the traffic,
  and show the effects of a condensation mechanism.
\end{abstract}

\section{A multi-speed exclusion process}\label{sec:multi}
The totally asymmetric exclusion process (\tasep) is a popular
statistical physics model of one-dimensional interacting particles
particularly adapted to traffic modeling. This is due to its simple
definition, and to the non-trivial exact solutions which have been
unveiled in the stationary regime~\cite{DeEvHaPa}. One important
shortcoming of this model is that it does not allow particles to move
at different speeds.  Cellular automata like the 
Nagel-Schreckenberg model~\cite{NaSch}
address this issue, leading to very
realistic though still simple simulators.  However, these 
models are difficult to handle mathematically beyond the mean field
approximation \cite{ChSaSc} and an approximate mapping with the asymmetric
chipping model suggests that the jamming phenomenon  takes place as a
broad crossover rather than a phase transition \cite{LeZi}. 
In this paper, we are interested in
analyzing the nature of fluctuations in the fundamental diagram
(\textsc{fd}), that is the mean flow of vehicles plotted against the
traffic density.  To address this question, we propose to extend the
\tasep in a different way, more convenient for the analysis albeit
less realistic from the point of view of traffic.

\begin{figure}
\begin{center}
\resizebox*{0.7\textwidth}{!}{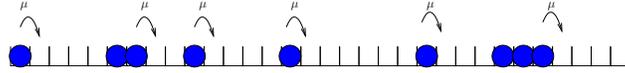}
\end{center}
\caption{The basic $1$-dimensional \tasep model.}\label{fig:tasep}
\end{figure}
The elementary $1$-dimensional \tasep (Fig.~\ref{fig:tasep}) model is
defined on a discrete lattice (e.g.\ a finite ring with boundary
periodic conditions, a segment with edges or an infinite line), where
each site may be occupied with at most one particle. Each particle
moves independently to the next site (say, to the right), at the times
of a Poisson process with intensity $\mu$. Therefore, the model is a
continuous time Markov process, which state is the binary encoded
sequence $\sigma_t \in \{O,V\}^N$ of size $N$ (the size of the system),
where the letter $V$ (resp.\ $O$) denotes a vehicle (resp.\ an empty space) at
site $i$. Each transition involves two consecutive letters when a particle moves from site $i$ to site $i+1$:
\[
VO\   \ra^{\mu}\  OV.
\]

\begin{figure}
\begin{center}
\scalebox{0.4}{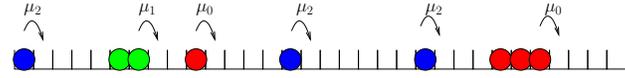}
\end{center}
\caption{The multi-speed $1$-dimensional \tasep model.}\label{fig:multi}
\end{figure}

In order to encode various speed levels, we propose to extend the
basic \tasep by allowing the particle to jump at different possible
rates which themselves vary stochastically in
time (Fig.~\ref{fig:multi}). Assuming for now a finite number of $n-1$
speed levels, the Markov chain that we consider is a sequence
encoded into a $n$-alphabet $\{O,A,B,\ldots\}$
\[
\sigma_t = \{V_i,\ldots,  i=1\ldots N\},\qquad V_i\in\{O,A,B,\ldots\}
\]
where $O$ is again an empty site, and $A$, $B,\ldots$ represent
occupied sites with jump rates $\mu_a, \mu_b, \ldots$. In our
model, the transitions remain local: the particle may jump to the next
site only if it is empty, we allow the final state to be conditioned
by the site after the next. More precisely, we assume that any
transition involves three consecutive letters, and distinguish between
two cases:
\[
\begin{cases}
\DD \ldots X O Y\ldots \ra^{\mu_x} \ldots O Y Y\ldots,\qquad Y\ne O\qquad\text{rule 1,}\\[0.2cm] 
\DD \ldots X O O\ldots \ra^{\mu_x} \ldots O Z O\ldots,\qquad Z\ne O\qquad\text{rule 2.}     
\end{cases}
\]
In the second case, the type (or equivalently the jump rate) of the
particle is chosen randomly according to a distribution $F$. 
As a limiting case, we will consider a
general continuous distribution $F$ on $\mathbb{R}^+$. In other words, a particle
at site $i$ with rate $\mu_x$ jumps to site $i+1$ and acquires a new
rate $\mu_z$ which is a random function of $V_{i+2}$.

The basic assumption is that if a car gets in close contact to another
one, it will adopt its rate. Conversely, if it arrives at a site not
in contact with any other car, the new rate will be freely determined
according to some random distribution. This models the acceleration or
deceleration process in an admittedly crude manner.  This setting is
different from usual exclusion processes with multi-type particles,
each having its own jump rate. It is more in line with the
Nagel-Schreckenberg model, with the difference that only local jumps
are allowed and speed is replaced by jump rate.

\section{$L$-stage tandem queue reformulation}\label{model}
In the context of exclusion processes, jams are represented as cluster of 
particles. Clustering phenomena can be analyzed
in some cases by mapping the process  to a tandem queueing network (i.e.\ a
zero range processes in statistical physics terms). For the simple
\tasep on a ring two dual mappings are possible:
\begin{itemize}
\item the queues are associated to empty sites and the clients are
the particles in contact behind this site, 
\item the queues are associated with particles and the clients are
the empty sites in front of this particle.   
\end{itemize}
By using one of these mappings, the \tasep is equivalent to a closed
cyclic queueing network, with fixed service rates equal to the jumping
rate $\mu$ of the particles. Steady states of such queueing network
have been analyzed thoroughly (see for ex.\ Kelly~\cite{Kel}) in terms
of a simple product form structure which we expose now.

Consider an open $L$-stage tandem queue, with arrival rate $\lambda$
and a common service rate $\mu$: $L$ queues with service rate $\mu$
are arranged in successive order (the departures from a given queue
coincide with the arrivals to the next one) and the arrival process of
the first queue is Poisson with intensity $\lambda$. Each queue is
stable when $\lambda<\mu$, transient when $\lambda>\mu$.   
It is then well known that the distribution of the number of clients
$X_1,\ldots,X_L$ in the queues is
\begin{equation}\label{eq:prod}
P(\{X_i=x_i,i=1\ldots L\}) = \prod_{i=1}^L P_\lambda(X_i=x_i),
\end{equation} 
where
\[
P_\lambda(X_i=x) = (1-\rho)\rho^x,\qquad\text{with}\qquad\rho \egaldef
\frac{\lambda}{\mu}.
\]

If the network is closed (the last queue is connected to the first one
in the ring geometry), then expression (\ref{eq:prod}) remains valid,
with the constraint that the total number of clients $\sum_{i=1}^L
X_i$ is fixed. In this case, $\lambda$ can be chosen arbitrarily, as
long as each queue in isolation remains ergodic.

It can be shown easily~\cite{PuMe} that, for a plain \tasep on a ring,
the size of the jams is asymptotically a geometric random variable
with parameter $\rho$ (when $N=kL$, $k=1,2\ldots$). In the open
geometry, the arrival rate is an external parameter which can be set
between $0$ and $\mu$. When it becomes comparable to the service rate,
i.e. when $\rho\simeq 1$, large queues may form and a random walk
first time return calculation yields a realistic scaling
behavior for the lifetime distribution of jams~\cite{NaPa}
\[
P(t)\simeq t^{-3/2}.
\]

\begin{figure}
\begin{center}
\scalebox{0.4}{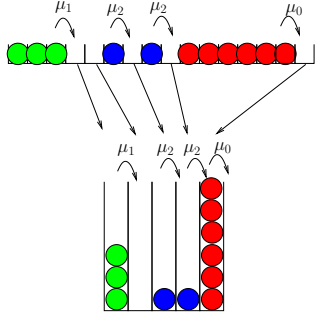}
\end{center}
\caption{Mapping of the variable-speed \tasep to a tandem
  queue}\label{fig:mapping}
\end{figure}

\bigskip In our multi-speed exclusion process, particles are guaranteed by
construction to form clusters with homogeneous speed, and the mapping
of empty sites to queues is suitable (Fig.~\ref{fig:mapping}). The new
feature is that the service rate $R_i$ of a given queue can change with
time: it is drawn randomly from a distribution with cumulative
distribution $F$ when the first customer arrives. It is assumed
that there exists a minimal service rate $\mu_0>0$ such that $F(\mu_0)=0$.
$\rho_0= \lambda/\mu_0 \le 1$ is therefore the
maximal possible load. The state is determined by the pair $(X,R)$ 
and the possible transitions are as follows:
\begin{align*}
(X=x,R=\mu)&\xrightarrow{\lambda\ind{x>0}} (X=x+1,R=\mu),\\
(X=0,R=\mu)&\xrightarrow{\lambda F(d\mu')} (X=1,R=\mu'),\\
(X=x,R=\mu)&\xrightarrow{\mu\ind{x>0}} (X=x-1,R=\mu).
\end{align*}

\begin{figure}
\begin{center}
\resizebox*{.8\textwidth}{!}{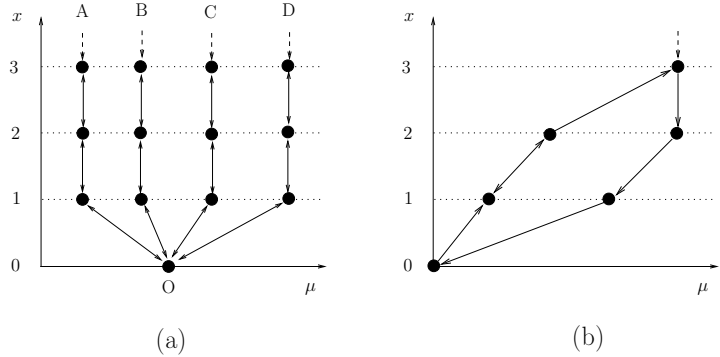}
\end{center}
\caption{State graph for isolated queues in the case of 
  (a) the multi-speed process definition, and (b) a non-reversible
  queue with hysteresis. States are represented by black dots and
  transitions by arrows.}\label{fig:stategraph}
\end{figure}

Since these transitions form a tree (see
Fig.~\ref{fig:stategraph}(a)), each queue in isolation is a reversible
Markov process and its stationary distribution reads:
\[
P_\lambda(X=x,R \in d\mu) = P_\lambda(X=0) 
F(d\mu)\left(\frac\lambda \mu\right)^x,
\]
with 
\[
P_\lambda(X=0) = \left(\int_{\mu_0}^\infty 
F(d\mu)\frac{\mu}{\mu-\lambda}\right)^{-1}.
\]

The distribution of the number of customers in the queue is therefore
no longer geometric:
\begin{equation}\label{eq:singlepdx}
  P_\lambda(X=x)
     =\int_{\mu_0}^\infty P_\lambda(X=x,R \in d\mu)
     =P_\lambda(X=0)\int_{\mu_0}^\infty F(d\mu)\left(\frac\lambda \mu\right)^x.
\end{equation}

Nevertheless, the product form expression (\ref{eq:prod}) for the
invariant measure remains valid, because of the reversibility of the
individual queues taken in isolation (see again~\cite{Kel}).  The
stationary distribution of the $L$-stage tandem queue takes the form
\[
P_\lambda(S) = \prod_{i=1}^L P_\lambda(X_i,R_i),
\]
for any sequence $S = \{(X_i,R_i), i=1,\ldots,L\}$. While each queue
has a different service rate at a given time, all the queues have
globally the same distribution. Our model is therefore encoded in the
single queue stationary distribution $P_\lambda(X_i,R_i)$.

\section{The fundamental diagram}
As announced in Section~\ref{sec:multi}, we turn now to the
fundamental diagram (\textsc{fd}), that is the plot of the mean flow
of vehicles against the traffic density. By nature, the fluctuations
in the \textsc{fd} are associated to the jam formation. Schematically,
three main distinct regimes or traffic phases have been identified by
empirical studies~\cite{Kerner}: one for to free-flow, and two
congested states, the ``synchronized flow'' and the ``wide moving
jam''.

In the case of the basic \tasep, it is well known, and rigorously
proved in some cases, that an hydrodynamic limit can be obtained by
rescaling both the spatial variable $x = i/N$ and the jumping rate
according to $\mu(N)= N V_0$, where $N$ is a rescaling which we let to
$\infty$ and $V_0$ is a constant. The corresponding coarse grained
density $\rho$ satisfies the inviscid Burger equation
\[
\frac{\partial\rho}{\partial t} = V_0\frac{\partial}{\partial x}\Bigl[ \rho(1-\rho)\Bigr].
\]
The \textsc{fd} at this scale is deterministic, since
\[
J(\rho) = V_0\rho(1-\rho),
\]
and symmetric w.r.t. $\rho=1/2$ because of the particle-hole symmetry.
$V_0$ represents the free velocity of cars, when the density is very low.

In practice, points plotted in experimental \textsc{fd} studies are
obtained by averaging data from static loop detectors over a few
minutes (see e.g.~\cite{Kerner}).  This is difficult to do with our
queue-based model, for which a space average is much easier to obtain.
The equivalence between time and space averaging is not an obvious
assumption, but since jams are moving, space and time correlations are
combined in some way~\cite{NaPa} and we consider this assumption to be
quite safe. In what follows, we will therefore compute the \textsc{fd}
by considering either the joint probability measure
$P_\lambda(d,\phi)$ for an open system, or the conditional probability
measure $P_\lambda(\phi\vert d)$ for a closed system, where
\[
\begin{cases}
\DD d = \frac{N}{N+L},\\[0.2cm]
\DD \phi = \frac{\Phi}{N+L},
\end{cases}
\ \text{with}\ 
\begin{cases}
L& \text{number of queues}\\
\DD N = \sum_{i=1}^L X_i&\text{number of vehicles}\\[0.2cm]
\DD \Phi = \sum_{i=1}^L R_i\ind{X_i>0}&\text{integrated flow}
\end{cases}
\]
are spatial averaged quantities and represent 
respectively the density and the traffic flow. 
We perform
the analysis in the ring geometry: this avoids edge effects, fixes
the numbers $N$ of vehicles and $L$ of queues, and finally makes sense
as an experimental setting. In the statistical physics parlance, the
fact that $N$ is fixed means that we are working with the canonical
ensemble. As a result this constraint yields the following form of the
joint probability measure:
\[
P(S) = \frac{1}{Z_L[N]} \prod_{i=1}^L P_\lambda (x_i,\mu_i),
\]
with the canonical partition function
\[
Z_L[N] \egaldef \sum_{\{x_i\}} \prod_{i=1}^L P_\lambda (x_i) 
\delta\bigl(\textstyle N-\sum_{i=1}^L x_i\bigr).
\]
These expressions are actually
independent of $\lambda$ in this specific ring geometry.
The density-flow conditional probability distribution 
takes the form
\begin{equation}\label{eq:fd}
P(\phi|d) =  \frac{Z_L[N,\Phi]}{Z_L[N]},
\end{equation}
with
\[
 N = \frac{d}{1-d}L,\qquad \Phi = \frac{\phi}{1-d}L,
\]
and
\begin{equation}\label{def:ZLNPhi}
Z_L[N,\Phi] \egaldef \sum_{\{x_i\}} \int\!\!\cdots\!\!\int\prod_{i=1}^L P_\lambda (x_i,d\mu_i) 
\textstyle \delta\bigl(N-\sum_{i=1}^L x_i\bigr)
\delta\bigl(\Phi-\sum_{i=1}^L \mu_i\ind{x_i>0}\bigr).
\end{equation}
Note (by simple inspection, see e.g. \cite{Kel})
that $P(\phi|d)$ is independent of $\lambda$.  

\section{Phase transition and condensation mechanism}
The connection between spontaneous formation of jams and the
Bose-Einstein condensation has been analyzed in some specific models,
with e.g.\ quenched disorder~\cite{Evans}, where particles are
distinguishable with different but fixed hopping rates attached to
them. In the present situation, all particles are
identical, but hopping rates may fluctuate, which is related to
annealed disorder in statistical physics.  The condensation mechanism
for zero range processes within the canonical ensemble has been
clarified in some recent work~\cite{EvMaZi}. Let us translate in our 
settings the main features of the condensation mechanism.  Assume that
the number of clients $X$ of an isolated queue has a long-tailed distribution
\[
P(X=x) \pro_{x\gg 1} \frac{1}{x^\alpha},\quad \alpha>1.
\]
The empirical mean queue size reads
\[
\bar X  = \frac{1}{L}\sum_{i=1}^L X_i,\qquad\text{and}\qquad
\EE \bar X = \EE_\lambda(X) 
 \egaldef \frac{\int_{\mu_0}^\infty F(d\mu)\frac{\lambda\mu}{(\mu-\lambda)^2}}
     {\int_{\mu_0}^\infty F(d\mu)\frac{\mu}{\mu-\lambda}},
\]
where $\EE_\lambda(X)$ is the expected number of clients in an isolated queue, 
when the arrival rate is $\lambda$.
Within the canonical ensemble, $\bar X$ is fixed, while for the grand
canonical ensemble, only the expectation $\EE(\bar X)$ is fixed.
In both cases, for $\alpha> 2$ there exists $\bar X_c$ such that,
when $\bar X>\bar X_c$ ($\EE(\bar X) >\bar X_c$ in the grand
canonical ensemble), one of the queues condenses, i.e.\ carries a
macroscopic number of particles. When $\alpha>2$, there is a
condensate with probability weight $O(L^{1-\alpha})$.

This condensation corresponds to a second order phase transition, and
occurs at a critical density $d_c$ which is the same in the canonical
and grand-canonical formalism. To determine $d_c$, first consider
\[
\bar d(\lambda) \egaldef \frac{\EE_\lambda(X)}{1+\EE_\lambda(X)}.
\]
$\EE_\lambda(X)$ increases monotonically with $\lambda$, which cannot
exceed $\mu_0$ (see Section~\ref{model}). Therefore, if
$\EE_{\mu_0}(X)\ =\ x_c\ <\ \infty$, then there exists a critical
density
\[
d_c = \frac{x_c}{1+x_c},
\]
such that for $d\ge d_c$ one of the queues condenses.
The interpretation is that $N_c  = L x_c$
is the maximal number of clients that can be in the queues in the fluid
regime, and the less costly way to absorb the excess $N-N_c$ is to put
it in one single queue.  Let us give an example, by specifying the
joint law through
\begin{equation}\label{def:vdist}
P(\mu_0\le R\le \mu_0y)= F(\mu_0 y)=
\Bigl(\frac{y-1}{r-1}\Bigr)^\alpha,\qquad 1\le y \le r,
\end{equation}
where $r>1$ is ratio between the highest and lowest speed.
In that case, using (\ref{eq:singlepdx}) we have the following
asymptotic as $\xi\to\infty$
\[
P_\lambda(X= x)  \propto\  \Bigl(\frac{\lambda}{\mu_0}\Bigr)^x
 \int_1^{r}(y-1)^{\alpha-1} y^{-x}dy 
\sim \frac{1}{x^{\alpha}} 
\Bigl(\frac{\lambda}{\mu_0}\Bigr)^x,
\]
and $\EE_{\mu_0}(X) <\infty$ when $\alpha>2$, which yields the 
possibility of condensation above the critical density
\begin{equation}\label{eq:dc}
d_c(\alpha,r) = \frac{(r-1)^{\alpha-2}}{\alpha-2+(r-1)^{\alpha-2}}.
\end{equation}

\section{Numerical results}
\begin{figure}[t]
\centering
\includegraphics*[width=0.49\textwidth]{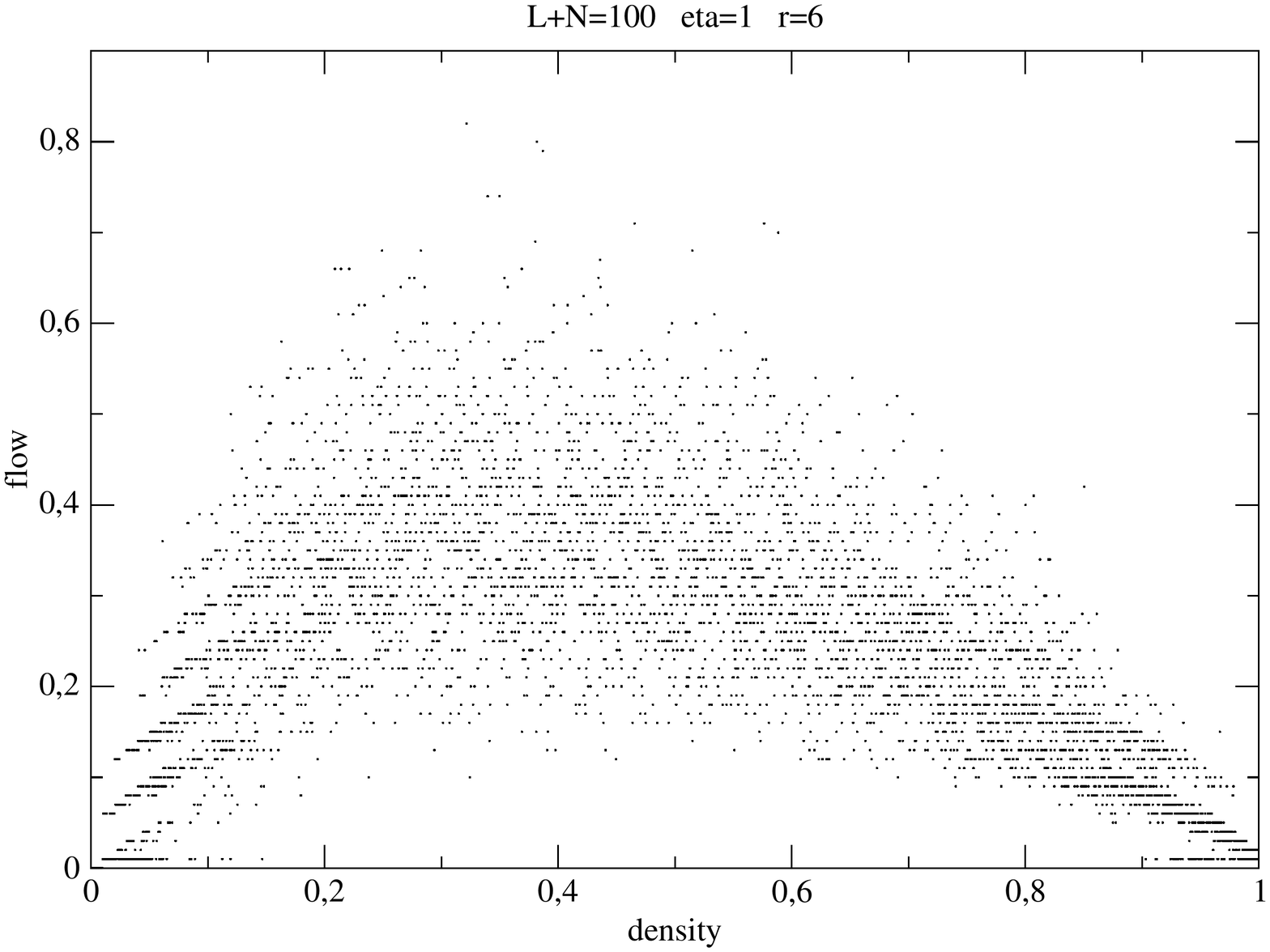}
\includegraphics*[width=0.49\textwidth]{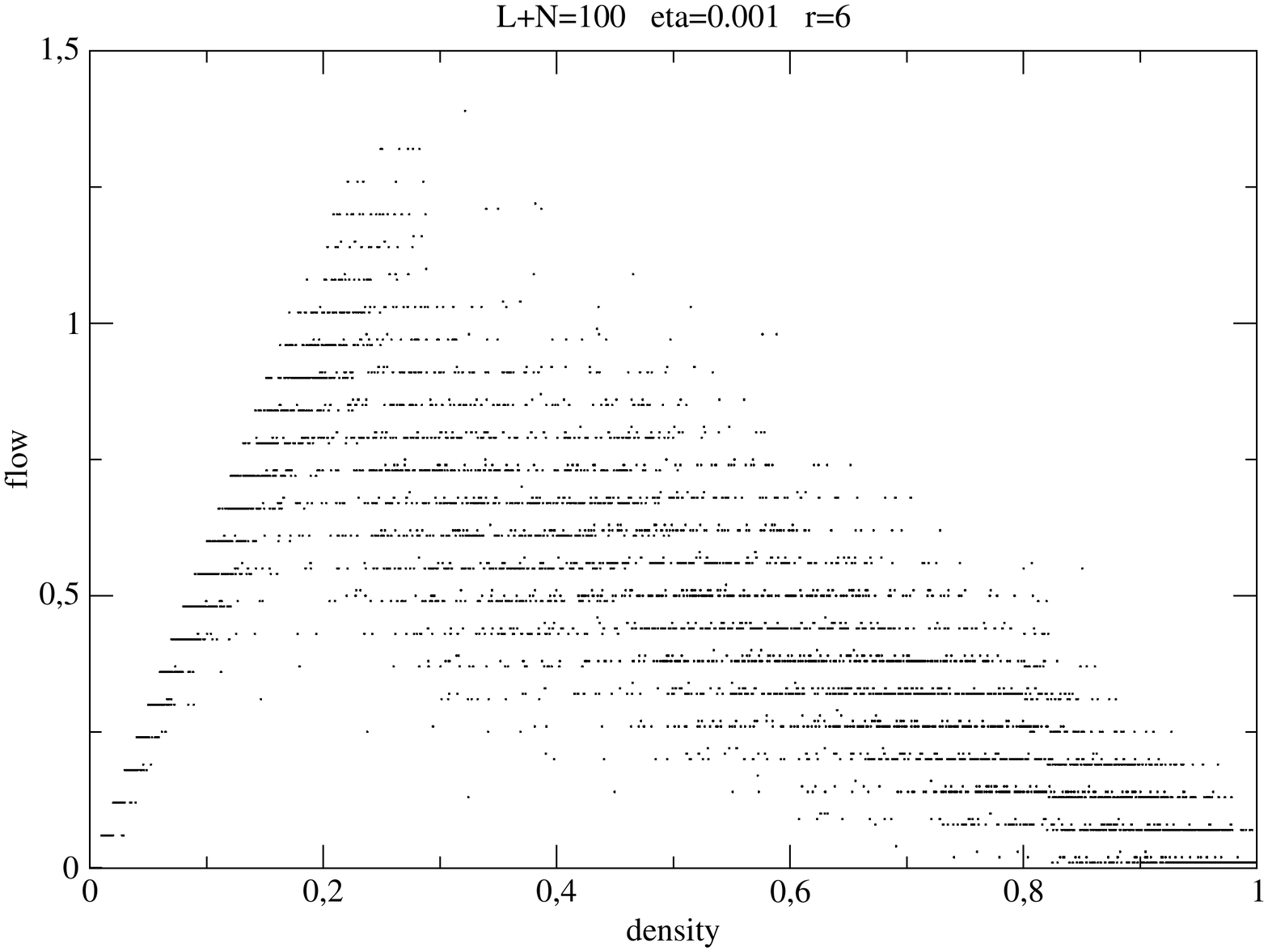}\\[0.2cm]
\hspace{-1cm}\includegraphics*[width=0.49\textwidth]{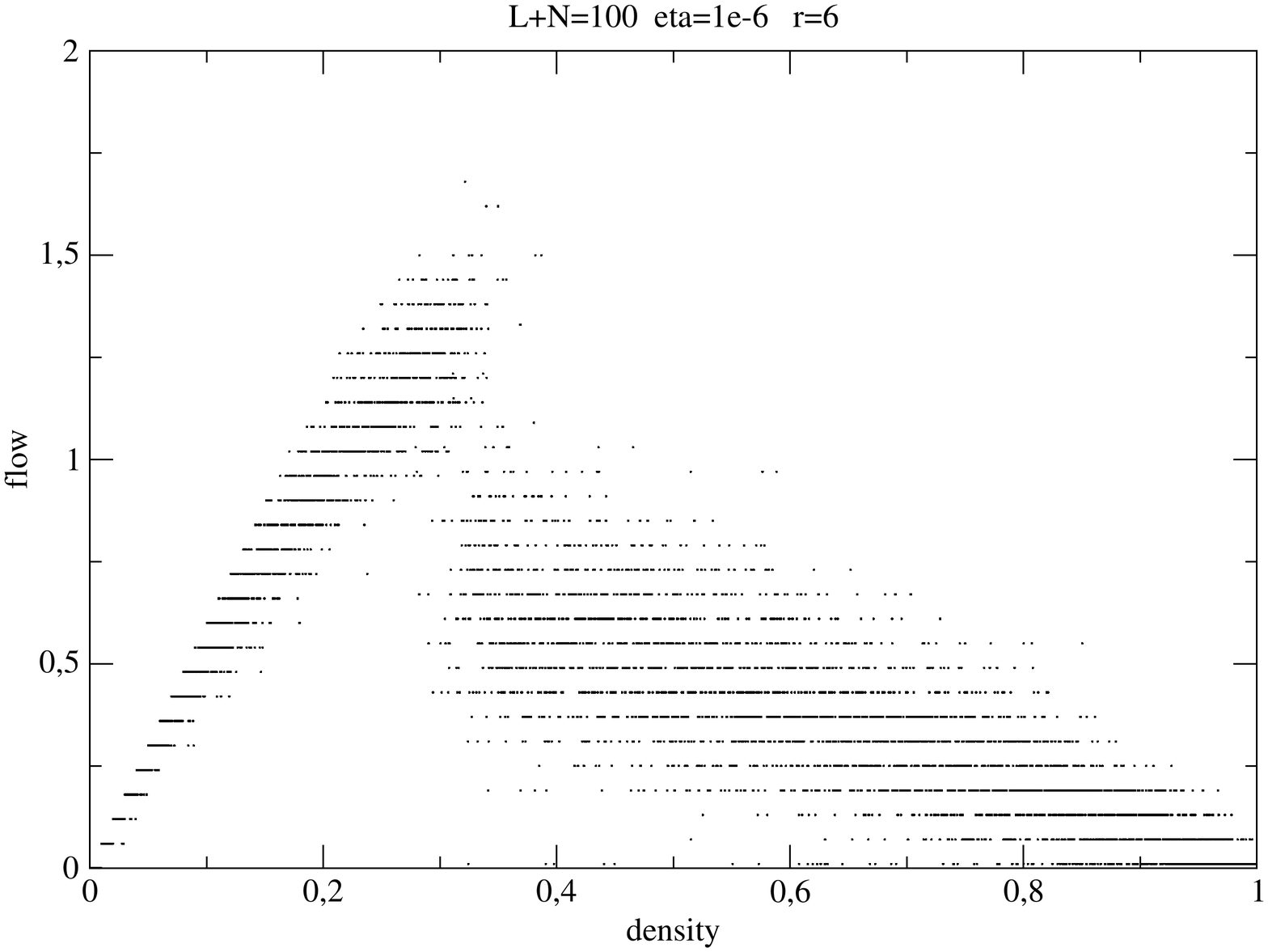}
\includegraphics*[width=0.41\textwidth]{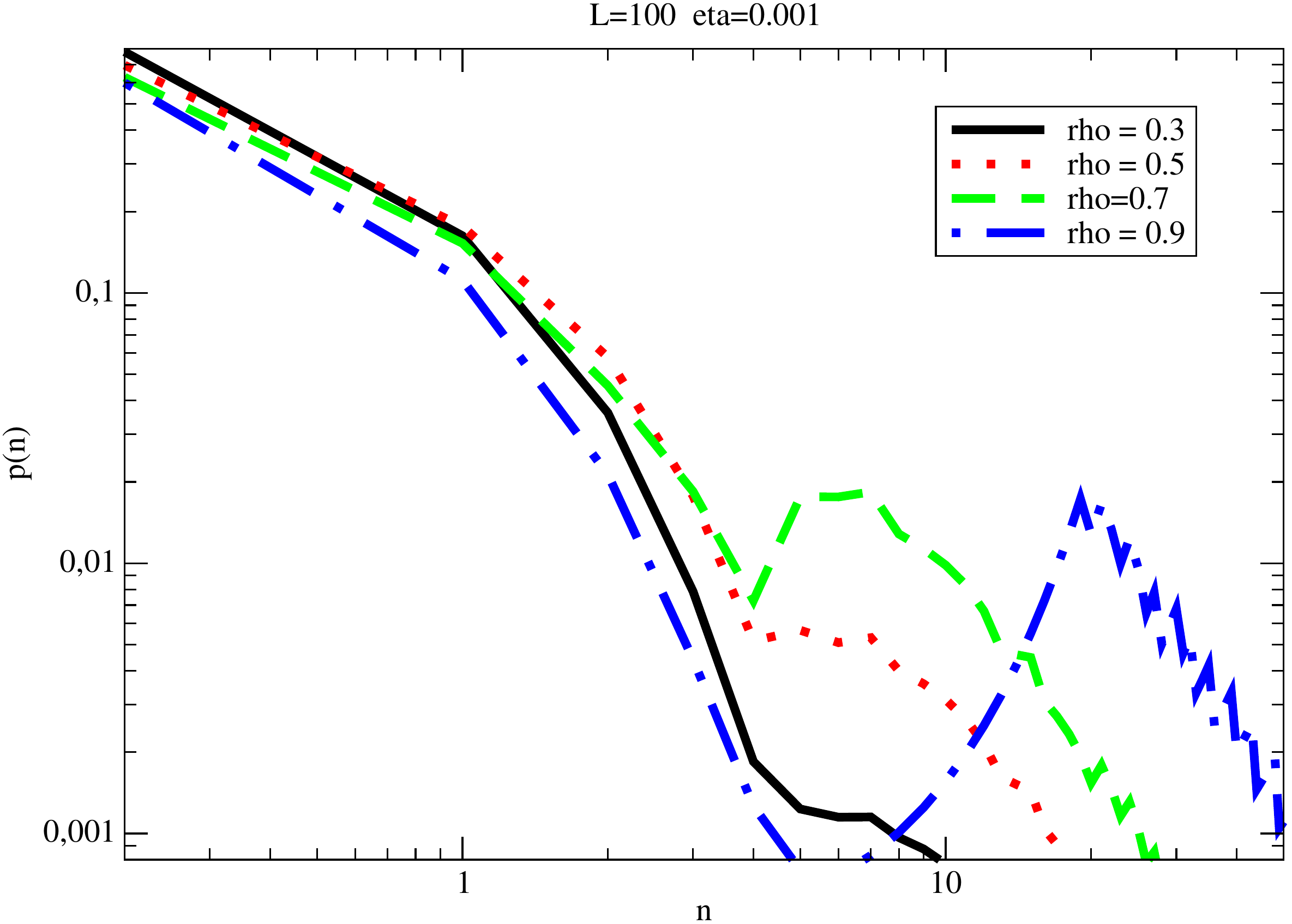}
\caption{Probabilistic fundamental diagram on the ring geometry
  ($L+N=100$) for a two-level speed distribution with $\eta=1$ (a),
  $0.001$ (b) and $10^{-6}$ (c); corresponding single queue
  distribution as a function of the density for $\eta=0.001$ (d).
}\label{fig:fd}
\end{figure}

\begin{figure}[t]
\centering
\includegraphics*[width=0.48\textwidth]{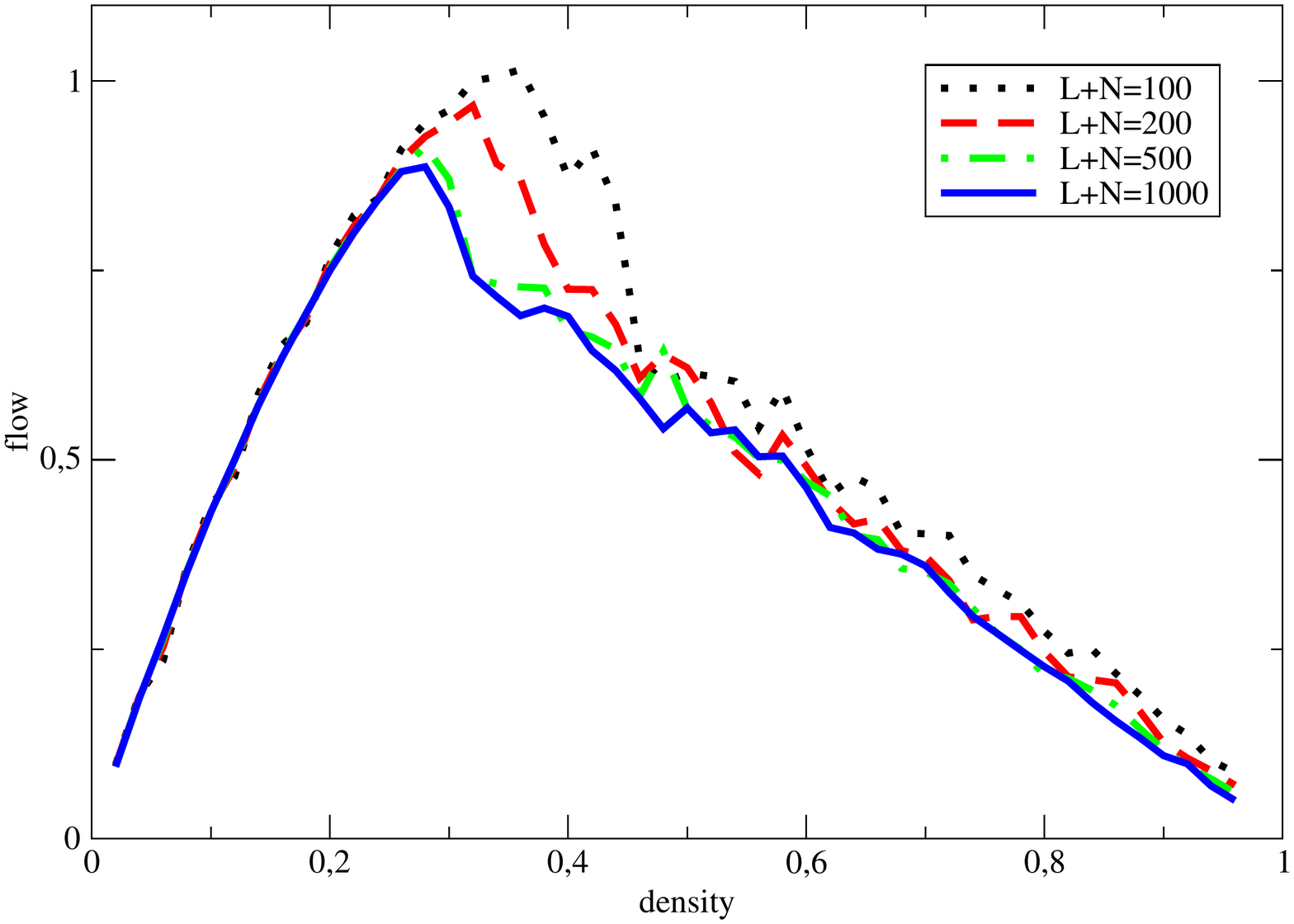}
\includegraphics*[width=0.48\textwidth]{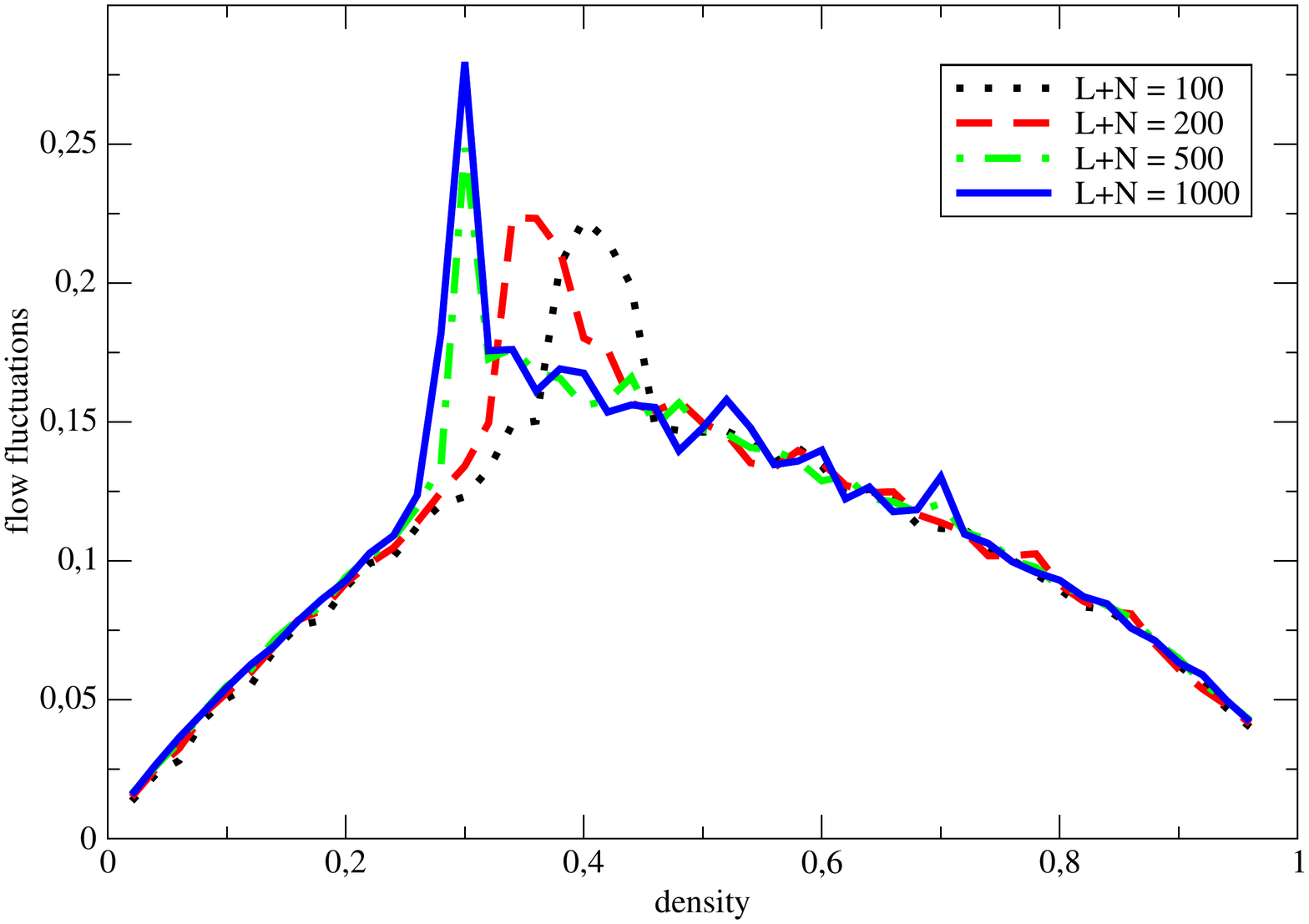}
\caption{Mean flow as a function of density for a
  continuous speed distribution ($\alpha=3$) on the ring geometry
  with varying sizes $L+N$ (left) and corresponding standard deviation
  rescaled by $\sqrt{L}$ (right). }\label{fig:fdalpha}
\end{figure}
The analysis of (\ref{eq:fd}) in the ring geometry can in principle be
performed by means of saddle point techniques~\cite{EvMaZi,FaLa}, which
we postpone to another work. Instead we present a numerical
approach: the \textsc{fd} presented in Fig.~\ref{fig:fd}(a)-(c) is
obtained by solving the recursive relation
\[
 Z_L[N,\Phi] = P_\lambda(X=0)Z_{L-1}[N,\Phi]
+ \sum_{x=1}^N \int P_\lambda(x,d\mu) 
Z_{L-1}[N-x,\Phi-\mu\ind{x>0}],
\]
up to some value $L_{MAX}=100$ for the number of queues, with a fixed
value of $\lambda<\mu_0$. Although in principle one arbitrary value of
$\lambda$ should suffice, in practice, the results for different
values of $\lambda$ have to be superposed in the diagram to get
significant results. Since this recursion is only tractable with a
finite number of possible velocities, the distribution $F$ used here is 
concentrated to two values $\mu_0$ and $\mu_1$. The presence of a
discontinuity in the fundamental diagram for small values of
$\eta\egaldef P(R=\mu_0)/P(R=\mu_1)$ is a finite size effect, which
disappears when the system size is increased while $\eta$ is kept
fixed. Nevertheless, the direct simulation of the closed $L$-stage
tandem queues, with continuous distribution (\ref{def:vdist}),
indicates as expected a second order phase transition when $\alpha>2$
(Fig.~\ref{fig:fdalpha}).  This transition is related to the formation
of a condensate, which is marked by the apparition of a bump in the
single queue distribution at the critical density (see
Fig.~\ref{fig:fd}(d)).  This condensation mechanism is responsible for
the slope discontinuity.  Fluctuations scale like $1/\sqrt{L}$, as
expected from the Central Limit Theorem. Note however that the
critical density is different than the one given by (\ref{eq:dc}) for
the open system.

\section{Perspectives}
In this work, we analyze the fluctuations in the fundamental diagram of
traffic by considering models from statistical physics and using
probabilistic tools. We propose a generalization of the \tasep
by considering a multi-speed exclusion process which is conveniently
mapped onto an $L$-stage tandem queue. When the individual queues are
reversible, general results from queueing network theory let us obtain
the exact form of the steady state distribution.  This measure is
conveniently shaped to compute the \textsc{fd}. Depending on
the speed distribution, it may present two phases, the free-flow and the
congested ones, separated by a second order phase transition. This
transition is associated to a condensation mechanism, 
when slow clusters are sufficiently rare. 

In practice, it is conjectured~\cite{Kerner} that there are three
phases in the \textsc{fd}, separated by first order phase transition.
A large number of possible extensions of our model are possible, by
playing with the definition of the state graph of a single queue
(Fig.~\ref{fig:stategraph}(a)). This graph accounts either for the
dynamics of single vehicle clusters, when queues are associated to
empty sites, or to the behavior of single drivers when queues are
associated to occupied sites.  In order to obtain first order phase
transitions, we will consider in future work models where the single
queues are not reversible in isolation, for example because of an
hysteresis phenomenon (Fig.~\ref{fig:stategraph}(b)).

\bibliography{refer}
\bibliographystyle{unsrt}

\end{document}